\documentclass[12pt]{article}
\usepackage{amsmath,amssymb}
\newcommand{\newsection}[1]{\section{#1}\setcounter{equation}{0}} 
\newcommand{\beq}{\begin{eqnarray}}
\newcommand{\eeq}{\end{eqnarray}}
\newcommand{\rd}{\partial}

\newcommand{\Tr}{\mathop{\mathrm{Tr}}}
\newcommand{\bbC}{\mathbb{C}}
\newcommand{\bbP}{\mathbb{P}}
\newcommand{\bbR}{\mathbb{R}}

\newcommand{\calE}{\mathcal{E}}
\newcommand{\calT}{\mathcal{T}}
\newcommand{\calU}{\mathcal{U}}
\newcommand{\calZ}{\mathcal{Z}}
\newcommand{\Gr}{\mathrm{Gr}}
\newcommand{\Fl}{\mathrm{Fl}}
\newcommand{\Pexp}{\mbox{\rm P-exp}}
\begin{document}

\title{Anti-self-dual Yang-Mills equations on 
noncommutative spacetime}
\author{Kanehisa Takasaki\\
{\normalsize Department of Fundamental Sciences, Kyoto University}\\
{\normalsize Yoshida, Sakyo-ku, Kyoto 606-8501, Japan}\\
{\normalsize E-mail: takasaki@math.h.kyoto-u.ac.jp}}
\date{}
\maketitle

\begin{abstract}
By replacing the ordinary product with the so called 
$\star$-product, one can construct an analogue of the 
anti-self-dual Yang-Mills (ASDYM) equations on the 
noncommutative $\bbR^4$.  Many properties of the ordinary 
ASDYM equations turn out to be inherited by the $\star$-product 
ASDYM equation.  In particular, the twistorial interpretation 
of the ordinary ASDYM equations can be extended to the 
noncommutative $\bbR^4$, from which one can also derive the 
fundamental structures for integrability such as a zero-curvature 
representation, an associated linear system, the Riemann-Hilbert 
problem, etc.  These properties are further preserved under 
dimensional reduction to the principal chiral field model and 
Hitchin's Higgs pair equations.  However, some structures 
relying on finite dimensional linear algebra break down in 
the $\star$-product analogues.  
\end{abstract}

\begin{flushleft}
\end{flushleft}
\bigskip

\begin{flushleft}
KUCP-153\\
{\tt hep-th/0005194}
\end{flushleft}


\newpage
\newsection{Introduction}

The deformed ADHM construction of 
Nekrasov and Schwarz \cite{bib:Ne-Sc} 
suggests that the anti-self-dual Yang-Mills (ASDYM) 
equations will be ``integrable'' on noncommutative 
spacetimes as well.  This is also advocated by 
the work of Kapustin et al.  \cite{bib:Ka-Ku-Or} 
that extends the ordinary twistorial interpretation 
of the ADHM construction \cite{bib:ADHM,bib:Atiyah} 
to the noncommutative $\bbR^4$.  Since twistor theory 
is a clue to the integrability of the ordinary ASDYM 
equations \cite{bib:Ma-Wo}, it is natural to expect 
that the ASDYM equations on the noncommutative $\bbR^4$, 
too, will be integrable.  

This issue is also interesting from the point of view 
of integrable systems of two-dimensional field theories, 
such as the principal chiral field (PCF) model 
\cite{bib:Polyakov} and Hitchin's Higgs pair equations 
\cite{bib:Hitchin-higgs}.  It is well known that 
these integrable systems can be derived from 
the ASDYM equations by dimensional reduction.  
If a similar reduction procedure works on the 
noncommutative $\bbR^4$, it seems likely that 
the integrability of the four-dimensional system 
will be inherited by the two-dimensional systems.  
This four-dimensional point of view can be an 
alternative approach to recent studies on the 
PCF and Wess-Zumino-Witten (WZW) models on 
noncommutative spacetimes 
\cite{bib:Mo-Sc,bib:Chu,bib:Da-Kr-La,bib:Fu-In,bib:Nu-Ol-Sc}.  

This paper aims to answer these questions.  
The gauge group is assumed to be $U(N)$ 
throughout the paper.  The ASDYM equations on 
the noncommutative $\bbR^4$ are then obtained 
from the ordinary ASDYM equations by replacing 
the product of fields in the field equations with 
the so called ``$\star$-product'' (the commutator 
of which is the Moyal bracket \cite{bib:Moyal}).  
We shall show that almost all part of the twistorial 
and integrable structures of the ordinary ASDYM 
equations can be extended to the noncommutative 
$\bbR^4$ by the same substitution rule.  What 
breaks down is the part where tools of finite 
dimensional linear algebra (determinants, 
Camer's formula, characteristic polynomials, 
etc.) are used.  

This paper is organized as follows.  Section 2 
presents the formulation of the $\star$-product 
ASDYM equations. Section 3 deals with the twistorial 
and integrable structures of the $\star$-product 
ASDYM equations.  Section 4 is concerned with 
some implications of the deformed ADHM construction. 
Section 5 is devoted to two-dimensional reductions.  
Section 6 is for conclusion.

\newsection{ASDYM equations on noncommutative $\bbR^4$}

\subsection{Spacetime coordinates}

The noncommutative $\bbR^4$ is characterized by the 
commutation relations 
\beq
    [x_j, x_k] = i \theta_{jk} 
\eeq
of the spacetime coordinates, where $\theta_{jk}$ are 
real constants.  These commutation relations can be 
extended to the associative $\star$-product 
\beq
    f \star g (x)
  = \left.\exp\left(\sum_{j,k=1}^4 
    \frac{i}{2}\theta_{jk}
    \rd_{x_j}\rd_{\tilde{x}_k}\right) 
    f(x) g(\tilde{x})\right|_{\tilde{x}=x} 
\eeq
of functions $f$ and $g$ on the spacetime.  

We now introduce complex coordinates $(z_1,z_2)$ that 
satisfy commutation relations of the form 
\beq
    [z_1, z_2] = - \zeta_{\bbC}, \quad 
    [\bar{z}_1,\bar{z}_2] = - \bar{\zeta}_{\bbC}, \quad 
    [z_1, \bar{z}_1] + [z_2, \bar{z}_2] = - \zeta_{\bbR}. 
\eeq
For instance, $z_1 = x_3 + i x_4$ and $z_2 = x_1 + ix_2$ 
give such complex coordinates after a suitable orthogonal 
transformation of the real coordinates. The complex 
constant $\zeta_\bbC$ and the real constant $\zeta_\bbR$ 
form a three-vector $(\zeta_{\bbR},\zeta_{\bbC})$ in 
$\bbR \times \bbC \simeq \bbR^3$, and can be rotated to 
any direction by the $SU(2)$ action 
\beq \label{eq:su2-action}
    \left(\begin{array}{cc}
      z_1 & z_2 \\
      - \bar{z}_2 & \bar{z}_1 
    \end{array}\right) 
    \quad \longmapsto \quad 
    \left(\begin{array}{cc} 
      \alpha & \beta \\
      - \bar{\beta} & \bar{\alpha} 
    \end{array}\right) 
    \left(\begin{array}{cc}
      z_1 & z_2 \\
      - \bar{z}_2 & \bar{z}_1 
    \end{array}\right) 
\eeq
($|\alpha|^2 + |\beta|^2 = 1$) on the spacetime 
coordinates.

\subsection{ASDYM equations}

Let $A$ be an $N \times N$ matrix-valued one-form 
representing a $U(N)$-connection.  Let 
$A_1, A_2, A_3, A_4$ denote the components in 
the real coordinate frame, and $A_{z_1}, A_{z_2}, 
A_{\bar{z}_1}, A_{\bar{z}_2}$ the components 
in the complex coordinate frame: 
\beq
    A = \sum_{j=1}^4 A_j dx_j 
      = \sum_{a=1,2} A_{z_a}dz_a  
      + \sum_{a=1,2} A_{\bar{z}_a}d\bar{z}_a. 
\eeq
The covariant derivatives can be accordingly written 
\beq
    \nabla_{{x}_j} = \rd_{x_j} + A_j, \quad 
    \nabla_{z_a} = \rd_{z_a} + A_{z_a}, \quad 
    \nabla_{\bar{z}_a} = \rd_{\bar{z}_a} + A_{\bar{z}_a}. 
\eeq
On the noncommutative $\bbR^4$, 
the components of the curvature two-form 
$F = \sum_{j,k} F_{jk} dx_j \wedge dx_k$ 
are defined as 
\beq
    F_{jk} = \rd_{x_j}A_k - \rd_{x_k}A_j 
           + [A_j, A_k]_\star. 
\eeq
Note that the usual matrix commutators 
$[A_j, A_k] = A_j A_k - A_k A_j$ are now replaced 
by the $\star$-product commutators 
\beq
    [A_j,A_k]_\star = A_j \star A_k - A_k \star A_j. 
\eeq
The components $F_{z_1 z_2}$, 
$F_{\bar{z}_1 \bar{z}_2}$ and $F_{z_a \bar{z}_b}$ 
in the complex coordinate frame are similarly 
written in terms of $A_{z_a}$ and $A_{\bar{z}_a}$. 
The ASDYM equations in the complex coordinate 
frame take the neat form 
\beq
    F_{z_1 z_2} = 0, \quad 
    F_{\bar{z}_1 \bar{z}_2} = 0, \quad 
    F_{z_1 \bar{z}_1} + F_{z_2 \bar{z}_2} = 0. 
\eeq

\subsection{Reduced form of ASDYM equations} 

It is well known that the ASDYM equations can be 
converted to a (classical) field theory with a 
Lagrangian formalism.  Actually, two types of 
such expressions are known.  One is Yang's 
equation \cite{bib:Yang} (also called the 
four-dimensional Donaldson-Nair-Schiff equation 
\cite{bib:Donaldson,bib:Na-Sc}).  Another 
expression is due to Leznov \cite{bib:Leznov} 
and Parkes \cite{bib:Parkes}. Both can be extended 
to the noncommutative spacetime as follows.   

Let us consider the first equation $F_{z_1 z_2} = 0$ 
of the ASDYM equations.  This is a partial 
(two dimensional) flatness condition.  
In the ordinary (complexified) spacetime, this 
implies that $A_{z_1}$ and and $A_{z_2}$ can be 
expressed as 
\beq
    A_{z_1} = h^{-1} \rd_{z_1}h, \quad 
    A_{z_2} = h^{-1} \rd_{z_2}h  
\eeq
with an $N \times N$ matrix-valued function $h$ 
of the spacetime coordinates.  (Of course, 
this is, in general, a local expression.)   
This expression persists to be true on the 
noncommutative spacetime if the ordinary product 
in the matrix multiplication are replaced by the 
$\star$-product: 
\beq
    A_{z_1} = (h)_\star^{-1} \star \rd_{z_1}h, \quad 
    A_{z_2} = (h)_\star^{-1} \star \rd_{z_2}h. 
\eeq
Here $(h)_\star^{-1}$ stands for an inverse 
with respect to the $\star$-product, namely, 
$h \star (h)_\star^{-1} = (h)_\star^{-1} \star h = 1$.  
One can prove this $\star$-product version of 
Frobenius' theorem in much the same way as a 
proof in the ordinary spacetime.  

In order to derive a $\star$-product analogue of 
Yang's equation, we solve another flatness condition 
$F_{\bar{z}_1,\bar{z}_2} = 0$ in the ASDYM equation as 
\beq
    A_{\bar{z}_1} = (k)_\star^{-1} \star \rd_{\bar{z}_1}k, 
    \quad 
    A_{\bar{z}_2} = (k)_\star^{-1} \star \rd_{\bar{z}_2}k, 
\eeq
and the ``matrix-ratio'' 
\beq
    g = k \star (h)_\star^{-1} 
\eeq
of $h$ and $k$.  As we shall show below, this matrix-valued 
field turns out to obey the field equation 
\beq
    \rd_{z_1}\Bigl((g)_\star^{-1} \star \rd_{\bar{z}_1} g\Bigr) 
  + \rd_{z_2}\Bigl((g)_\star^{-1} \star \rd_{\bar{z}_2} g\Bigr) = 0 
\eeq
This gives an analogue of Yang's equation on the 
noncommutative spacetime.  The Lagrangian formalism 
in the commutative case can be readily extended to 
the noncommutative case.  Note that the field equation 
\beq
    \rd_z\Bigl((g)_\star^{-1} \star \rd_{\bar{z}} g\Bigr) 
  + \rd_{\bar{z}}\Bigl((g)_\star^{-1} \star \rd_z g\Bigr) = 0 
\eeq
of the noncommutative PCF model \cite{bib:Da-Kr-La} 
can be derived by dimensional reduction.  

The foregoing noncommutative analogue of Yang's equation 
can be derived by the following trick.  Let us consider 
the finite gauge transformation by $h$.  Two of the four 
gauge potentials, $A_{z_a}$ ($a = 1,2$), are thereby 
gauged away as
\beq
    \nabla_{z_a} \to 
    h \circ \nabla_{z_a} \circ (h)_\star^{-1} 
    = \rd_{z_a}, 
\eeq
and the other two are transformed as 
\beq
    \nabla_{\bar{z}_a} 
    &\to& 
    h \circ \nabla_{\bar{z}_a} \circ (h)_\star^{-1} 
    \nonumber \\
    &=& \rd_{\bar{z}_a} - \rd_{\bar{z}_a}h\star(h)_\star^{-1} 
    + h \star (k)_\star^{-1}\star\rd_{\bar{z}_a}k\star(h)_\star^{-1}
    \nonumber \\
    &=& \rd_{\bar{z}_a} + (g)_\star^{-1}\star\rd_{\bar{z}_a}g. 
\eeq
The gauge potentials are now in a {\it half-flat} gauge 
in which two of the gauge potentials vanish, 
\beq
    A_{z_1} = 0, \quad A_{z_2} = 0, 
\eeq
and the other two gauge potentials are written as 
\beq
    A_{\bar{z}_a} = (g)_\star^{-1}\star\rd_{\bar{z}_a}g. 
\eeq
The remaining equation $F_{z_1\bar{z}_1} + F_{z_2\bar{z}_2} = 0$ 
of the $\star$-product ASDYM equations, which now takes the 
simplified form 
\beq
    \rd_{z_1}A_{\bar{z}_1} + \rd_{z_2}A_{\bar{z}_2} = 0, 
\eeq
gives the $\star$-product analogue of Yang's equation.   

One can see, from this derivation of Yang's equation, 
the existence of a field theoretical ``dual'' of Yang's 
equation as well.  Note that the $\star$-product ASDYM 
equations in the foregoing half-flat gauge with 
$A_{z_1} = A_{z_2} = 0$ consist of the two equations 
\beq
    \rd_{\bar{z}_1}A_{\bar{z}_2} 
    - \rd_{\bar{z}_2}A_{\bar{z}_1} 
    + [A_{\bar{z}_1}, A_{\bar{z}_2}]_\star = 0, 
    \quad 
    \rd_{z_1}A_{\bar{z}_1} + \rd_{z_2}A_{\bar{z}_2} = 0.  
\eeq
If one solves the first equation as a partial flatness 
condition, as we have seen above, Yang's equation 
emerges from the second equation.  Meanwhile, one can also 
solves the second equation as 
\beq
    A_{\bar{z}_1} = - \rd_{z_2} \phi, \quad 
    A_{\bar{z}_2} = \rd_{z_1} \phi 
\eeq
for a matrix-valued potential $\phi$.  The first equation 
then takes the form 
\beq
    (\rd_{z_1} \rd_{\bar{z}_1} + \rd_{z_2}\rd_{\bar{z}_2}) \phi 
    + [\rd_{z_1}\phi, \rd_{z_2}\phi]_\star = 0.  
\eeq
This is a $\star$-product version of the field equation 
of Leznov and Parkes.

\newsection{Twistor theory and integrability}

\subsection{Twistor geometry} 

Twistor theory encodes various fields on 
spacetime to a geometric structure on another 
(complex) manifold called the ``twistor space'' 
\cite{bib:Pe-Ri,bib:Wa-We}.   
In the case of four dimensional flat spacetime, 
the twistor space is the three dimensional complex 
projective space $\bbP_\bbC^3$.  Roughtly speaking, 
twistor theory is a kind of ``tomography'', namely, 
to ``scan'' the spacetime by a three-parameter 
family of two-dimensional surfaces (``twistor 
surfaces'') $S(\xi)$ labelled by the point $\xi$ 
of the twistor space.  We review the essence 
of twistor geometry in the following.  

To define the twistor surfaces, however, the real 
(Euclidean) spacetime $\bbR^4$ has to be extended 
to the complexified spacetime $\bbC^4$, in which 
$(z_1,z_2,\bar{z}_1,\bar{z}_2)$ are {\it independent} 
complex coordinates.  The twistor surfaces in the 
complexified spacetime $\bbC^4$ are labelled 
by three parameters $(\lambda,u_1,u_2)$, and 
defined by the equations 
\beq
    z_1 - \lambda \bar{z}_2 = u_1, \quad 
    z_2 + \lambda \bar{z}_1 = u_2. 
\eeq
The parameters $(\lambda,u_1,u_2)$ are local 
coordinates on a coordinate patch of the whole 
twistor space $\bbP_\bbC^3$. Furthermore, $\lambda$ 
turns out to play the role of the ``spectral parameter'' 
in the theory of integrable systems.  

Various real spacetimes, such as the Minkowski spacetime 
and the spacetime with $(2,2)$ signature, are embedded 
in the complexified spacetime $\bbC^4$ as ``real slices''. 
Although the twistor surface $S(\lambda,u_1,u_2)$ 
intersects with the Euclidean spacetime at most at 
a point, the intersection with the Minkowski spacetime 
is a null line, and the intersection with the $2+2$ 
spacetime is a totally null surface (i.e., the inner 
product of any two tangent vectors vanish).  

The twistor space $\bbP_\bbC^3$ itself appears 
in the description of a compacitified spacetime, 
such as the one-point compactification $S^4 = 
\bbR^4 \cup \{\infty\}$ of the Euclidean spacetime. 
Let us introduce the complex Grassmann variety 
\beq
    \Gr_\bbC(2,4) = \{ V_2 \mid V_2 \subset \bbC^4, \ 
    \dim V_2 = 2 \}
\eeq
of vector subspaces of $\bbC^4$ and the flag variety 
\beq
   \Fl_\bbC(1,2,4) = \{ (V_1,V_2) \mid 
    V_1 \subset V_2 \subset \bbC^4, \ 
    \dim V_1 = 1, \ \dim V_2 = 2 \}
\eeq
of pairs of nested vector subspaces of $\bbC^4$. 
The Grassmann variety is a natural complexification 
of $S^4$.  Twistor geometry connects these compact 
(and complexified) spacetimes with the twistor space 
$\bbP_\bbC^3$ by the ``Klein correspondence'' 
\beq
    \Gr_\bbC(2,4) \ 
    \stackrel{p_2}{\longleftarrow} \ 
    \Fl_\bbC(1,2,4) \ 
    \stackrel{p_1}{\longrightarrow} \ 
    \bbP_\bbC^3, 
\eeq
where the projections $p_1$ and $p_2$ send 
the flag $(V_1,V_2)$ to $V_1 \in \bbP_\bbC^3$ 
and $V_2 \in \Gr_\bbC(2,4)$, respectively.  
The subset $S(\xi) = p_2(p_1^{-1}(\xi))$ is 
isomorphic to $\bbP_C^2$ and gives a compactification 
of the foregoing twistor surface $S(\lambda,u_1,u_2)$ 
in $\bbC^4$.  Similarly, the subset $L(x) = 
p_1(p_2^{-1}(x))$ is isomorphic to $\bbP_\bbC^1$ 
and plays a key role in {\it decoding} the twistorial 
data. 

The uncompactified spacetime $\bbR^4$ (or, 
rather, its complexification $\bbC^4$) 
can be described by the open twistor space 
\beq
    \calT = \bbP_\bbC^3 \setminus \bbP_\bbC^1
\eeq
with a line $\bbP_\bbC^1$ deleted.  It is rather 
this twistor space that we mostly consider in 
the following.  This open twistor space has the 
projection 
\beq
    \begin{array}{rcl} 
    \pi: \calT &\longrightarrow& \bbP_\bbC^1 \\
    \xi = [\xi_0:\xi_1:\xi_2:\xi_3] &\longmapsto& 
    [\xi_0:\xi_1] 
    \end{array}
\eeq
and is covered by the two standard coordinate 
patches $\calU = \{\xi_0 \not= 0\}$ and $\hat{\calU} 
= \{\xi_1 \not= 0\}$.  The deleted line $\bbP_\bbC^1$ 
is the locus where $\xi_0 = \xi_1 = 0$.  The three 
parameters $(\lambda,u_1,u_2)$ can be identified with 
the standard local coordinates on $\calU$: 
\beq
    \lambda = \xi_1/\xi_0, \quad 
    u_1 = \xi_2/\xi_0, \quad 
    u_2 = \xi_3/\xi_0. 
\eeq
Thus, in particular, $\lambda$ is an affine coordinate 
of the base, and  $u_1$ and $u_2$ are coordinates along 
the fibers.

\subsection{Flatness on twistor surfaces} 

The three members of the ASDYM equations can be 
combined to a single equation of the form 
\beq
    F(\rd_{\bar{z}_1} - \lambda \rd_{z_2}, 
      \rd_{\bar{z}_2} + \lambda \rd_{z_1}) 
  = F_{z_1 z_2} 
    - \lambda(F_{z_1\bar{z}_1} + F_{z_2\bar{z}_2}) 
    + \lambda^2 F_{\bar{z}_1 \bar{z}_2} 
  = 0. 
\eeq
Here $F(v,v')$ stands for the contraction of $F$ 
by two vector fields $v,v'$.  Since the two 
vector fields $\rd_{\bar{z}_1} - \lambda \rd_{z_2}, 
\rd_{\bar{z}_2} + \lambda \rd_{z_1}$ on the 
left hand side span the tangent planes of the 
twistor surface $S(\lambda,u_1,u_2)$, the 
foregoing equation means the flatness 
\beq
    F|_{S(\lambda,u_1,u_2)} = 0 
\eeq
of the connection on all twistor surfaces.  

Frobenius' theorem connects this flatness 
(or ``zero-curvature'') condition with the 
integrability of the linear system 
\cite{bib:Be-Za,bib:Pohlmeyer,bib:Ch-Pr-Si} 
\beq \label{eq:diffeq-Psi}
    (\nabla_{\bar{z}_1} - \lambda\nabla_{z_2})
    \Psi(\lambda) = 0, 
    \quad 
    (\nabla_{\bar{z}_2} + \lambda\nabla_{z_1})
    \Psi(\lambda) = 0,  
\eeq
where $\Psi(\lambda)$ is a vector- or matrix-valued
unknown function (which, of course, depends on the 
spacetime coordinates as well).  Having this linear 
system, one can now apply a number of techniques for 
integrable systems to the ASDYM equations \cite{bib:Ma-Wo}.

Note that the first two equations of the ASDYM 
equations (from which $h$ and $k$ were derived) 
correspond to the flatness on the twistor surfaces 
with $\lambda = 0$ and $\lambda = \infty$.  
Accordingly, one can choose two matrix-valued 
solutions $\Psi(\lambda)$ and $\hat{\Psi}(\lambda)$ 
of 
(\ref{eq:diffeq-Psi}) to be such that 
\beq
    \Psi(0) = h, \quad 
    \hat{\Psi}(\infty) = k. 
\eeq
In other words, $\Psi(\lambda)$ and 
$\hat{\Psi}(\lambda)$ are one-parameter 
deformations of $h$ and $k$.  Moreover, 
the Laurent expansion 
\beq
    \Psi(\lambda) 
      = h + \sum_{n=1}^\infty w_n \lambda^n, 
    \quad 
    \hat{\Psi}(\lambda) 
      = k + \sum_{n=1}^\infty \hat{w}_n \lambda^{-n} 
\eeq
of these solutions of (\ref{eq:diffeq-Psi}) 
are related to two infinite series of nonlocal 
conservation laws \cite{bib:Pohlmeyer,bib:Ch-Pr-Si}.  
There is no reason that these two solutions 
coincide.  They rather give a pair that arise 
in the so called Riemann-Hilbert problem.  
We now turn to this issue.

\subsection{Vector bundle and Riemann-Hilbert problem}

The twistor transformation \cite{bib:Ward,
bib:At-Wa,bib:At-Hi-Si,bib:Co-Fa-Ya-Go} 
encodes a solution of the ASDYM equations to 
a holomorphic vector bundle $\calE$ over the 
twistor space.  Given a solution of the ASDYM 
equations, one can consider an associated rank $N$ 
vector bundle $E$ over the spacetime with an induced 
connection.  This connection is flat on each twistor 
surface $S(\xi)$. The fiber $\calE_\xi$ of the bundle 
$\calE$ at a point $\xi$ of the twistor space is, 
by definition, the vector space of flat sections of 
$E|_{S(\xi)}$. In a down-to-earth language, the fiber 
$\calE_\xi$ is the vector space of $E$-valued 
solutions of linear system (\ref{eq:diffeq-Psi}) 
restricted to $S(\lambda,u_1,u_2)$. 

This bundle $\calE$ need not be defined over the 
whole twistor space.  If the solution of the ASDYM 
equations is defined in a small neighborhood of 
a spacetime point $x$, the bundle $\calE$ is 
accordingly defined only in a neighborhood of 
the line $L(x) = \{\xi \in \bbP_\bbC^3 \mid x 
\in S(\xi)\}$.  Instanton solutions are global 
solutions that give rise to a globally defined 
vector bundle on the whole twistor space.  

The holomorphic vector bundle $\calE$ has 
the special property that the restriction 
$\calE|_{L(x)}$ to the line $L(x) \simeq 
\bbP_\bbC^1$ is holomorphically trivial for 
any spacetime point $x$ in the domain where  
the gauge potentials are defined. This 
property of $\calE$ plays a key role in the 
{\it inverse} transformation, namely, to 
reproduce the solution of the ASDYM equation 
from the vector bundle $\calE$.  

It is here that the notion of Riemann-Hilbert 
problem emerges.  Let us recall that any 
holomorphic vector bundle over $\bbP_\bbC^1$ 
can be represented by the ``patching function'' 
$p(\lambda)$ on the intersection $D \cap \hat{D}$ 
of two affine coordinate patches $\{D, \hat{D}\}$ 
of $\bbP_\bbC^1$.  The patching function $p(\lambda)$ 
is an $GL(N,\bbC)$-valued holomorpihc function.  
If the vector bundle is holomorphically trivial, 
the patching function can be expressed as 
\beq 
    p(\lambda) = \hat{\Psi}(\lambda)^{-1}\Psi(\lambda), 
\eeq
where $\Psi(\lambda)$ and $\hat{\Psi}(\lambda)$ 
are $GL(N,\bbC)$-valued holomorphic functions 
of $\lambda$ on $D$ and $\hat{D}$, respectively.  
Finding such a pair of matrix-valued functions 
to the given data $p(\lambda)$ is a kind of 
Riemann-Hilbert problem (also called  the 
``splitting'' problem in the terminology of Ward).  

The patching function $p(\lambda)$ is determined 
by a patching function of the vector bundle $\calE$ 
itself.  As already remarked, the twistor space 
$\calT = \bbP_\bbC^3 \setminus \bbP_\bbC^1$ is 
covered by the two coordinate patches $\calU$ 
and $\hat{\calU}$.  The vector bundle $\calE$ 
is described by a $GL(N,\bbC)$-valued function 
$P(\lambda,u_1,u_2)$ that glues together 
the rank-$N$ trivial bundles over $\calU$ and 
$\hat{\calU}$. Its restriction on the line $L(x)$ 
is nothing but the patching function $p(\lambda)$ 
of  $\calE|_{L(x)}$: 
\beq
    p(\lambda) = P(\lambda, z_1 - \lambda \bar{z}_2, 
                   z_2 + \lambda \bar{z}_1). 
\eeq
In particular, the patching function $p(\lambda)$ 
turns out to obey the linear differential equations 
\beq \label{eq:diffeq-p}
    (\rd_{\bar{z}_1} - \lambda \rd_{z_2})p(\lambda) = 0, 
    \quad 
    (\rd_{\bar{z}_2} + \lambda \rd_{z_1})p(\lambda) = 0. 
\eeq
(Note that this is a rather simplified setup.  
If the solution is defined in a general domain of 
spacetime, we need a more refined cohomological 
language --- see Ivanova's review \cite{bib:Ivanova}
and references cited therein.)

Given such a patching function, one can prove that 
the Riemann-Hilbert problem indeed solves the ASDYM 
equations.  We shall review this proof later on 
in the framework of the noncommutative spacetime. 
The converse is also true. Namely, if $\Psi(\lambda)$ 
and $\hat{\Psi}(\lambda)$ are a pair of {\it arbitrary} 
solutions of (\ref{eq:diffeq-Psi}), its matrix ratio 
$\hat{\Psi}(\lambda)^{-1}\Psi(\lambda)$ satisfies 
equation (\ref{eq:diffeq-p}). This can be confirmed 
by direct calculations.  

Solving the Riemann-Hilbert problem {\it explicitly} 
is usually very difficult.  Explicit solutions 
are known for special cases only.   The so called 
``Ward Ansatz'' (or ``Atiyah-Ward Ansatz'') solutions 
\cite{bib:At-Wa,bib:Ward-ansatz,bib:Iv-Ma-Ne} 
provide such an example.  The corresponding 
Riemann-Hilbert problem can be solved by linear 
algebra. 

The existence of a large set of hidden symmetries 
of the ASDYM equations \cite{bib:Ch-Ge-Wu,bib:Dolan} 
can be explained by the Riemann-Hilbert problem 
\cite{bib:Ue-Na,bib:Wu}.  Those symmetries are 
generated by the left and right action

\beq
    p(\lambda) \ \mapsto \ 
    g_L(\lambda) p(\lambda) g_R(\lambda)^{-1} 
\eeq
of $GL(N,\bbC)$-valued functions 
$g_L(\lambda)$ and $g_R(\lambda)$ of 
$(\lambda, z_1 - \lambda \bar{z}_2, 
z_2 + \lambda \bar{z}_1)$.   The infinitesimal 
form of these symmetries can be determined 
explicitly and reproduces the previously known 
results \cite{bib:Takasaki}.  For subsequent 
progress on finite transformations, see 
Popov's paper \cite{bib:Popov}.

\subsection{Integrability of $\star$-product ASDYM equations}

Having reviewed the twistorial and integrable 
structures of the ordinary ASDYM equations, 
we now turn to the $\star$-product ASDYM 
equations.  

The geometric setup of twistor theory can be 
extended to the noncommutative spacetime 
rather straightforward. To see this, 
let us notice that the commutation relations 
of the complex coordinates $(z_1, z_2, 
\bar{z}_1, \bar{z}_2)$ can be rewritten 
\beq
    [z_1 - \lambda\bar{z}_2, z_2 + \lambda\bar{z}_1] 
    = - \zeta_\bbC - \lambda\zeta_\bbR 
    + \lambda^2\bar{\zeta}_\bbC.  
\eeq
The linear combinations of the spacetime coordinates 
on the left hand side are exactly those in the 
definition of the twistor surface $S(\lambda,u_1,u_2)$. 
Accordingly, whereas $\lambda$ persists to be a 
commutative coordinate, the coordinates $u_1$ and 
$u_2$ of the fibers of $\pi:\calT \to \bbP_\bbC^1$ 
turn out to have to obey the commutation relation 
\beq
    [u_1, u_2] =  - \zeta_\bbC - \lambda\zeta_\bbR 
    + \lambda^2\bar{\zeta}_\bbC. 
\eeq
Thus the twistor space, like the spacetime, becomes 
a noncommutative manifold.  This will be an alternative 
interpretation of the results of of Kapustin et al. 
\cite{bib:Ka-Ku-Or}.  

The linear system for $\Psi(\lambda)$ is now 
replaced by the $\star$-product version 
\beq \label{eq:diffeq-Psi-star}
    (\nabla_{\bar{z}_1} - \lambda \nabla_{z_2})
    \star\Psi(\lambda)  
    = (\rd_{\bar{z}_1} - \lambda \rd_{z_2})\Psi(\lambda) 
    + (A_{\bar{z}_1} - \lambda A_{z_2})\star\Psi(\lambda) 
    = 0, 
    \nonumber \\
    (\nabla_{\bar{z}_2} + \lambda \nabla_{z_1})
    \star\Psi(\lambda) 
    = (\rd_{\bar{z}_2} + \lambda \nabla_{z_1})\Psi(\lambda) 
    + (A_{\bar{z}_2} + \lambda A_{z_1})\star\Psi(\lambda) 
    = 0. 
\eeq

Although the notion of vector bundles on the 
noncommutative twistor space is complicated 
\cite{bib:Ka-Ku-Or}, the Riemann-Hilbert problem 
itself remains intact except that the product 
$\hat{\Psi}(\lambda)^{-1} \Psi(\lambda)$ is 
replaced by the $\star$-product: 
\beq
    p(\lambda) = 
    (\hat{\Psi}(\lambda))_\star^{-1} \star \Psi(\lambda). 
\eeq
The patching function $p(\lambda)$ is required to 
satisfy the same linear differential equations 
as (\ref{eq:diffeq-p}), or, equivalently, to be 
of the form $P(\lambda,z_1 - \lambda\bar{z}_2,
z_2 + \lambda\bar{z}_1)$.  

Let us confirm that this Riemann-Hilbert problem 
indeed solves the $\star$-product ASDYM equations.  
The reasoning is fully parallel to the ordinary 
ASDYM equations.  We first note that 
(\ref{eq:diffeq-p}) implies the equations 
\beq
    (\rd_{\bar{z}_1} - \lambda \rd_{z_2}) 
    \Psi(\lambda) \star (\Psi(\lambda))_\star^{-1} 
    &=& 
    (\rd_{\bar{z}_1} - \lambda \rd_{z_2}) 
    \hat{\Psi}(\lambda) \star (\hat{\Psi}(\lambda))_\star^{-1}, 
    \nonumber \\
    (\rd_{\bar{z}_2} + \lambda \rd_{z_1}) 
    \Psi(\lambda) \star (\Psi(\lambda))_\star^{-1} 
    &=& 
    (\rd_{\bar{z}_2} + \lambda \rd_{z_1}) 
    \hat{\Psi}(\lambda) \star (\hat{\Psi}(\lambda))_\star^{-1}. 
\eeq
Since $D$ and $\hat{D}$ cover the whole Riemann 
sphere, both hand sides of these equations define 
a matrix-valued meromorphic function with the only 
possible poles being at $\lambda = \infty$ and of 
the first order.  By Liouville's theorem, 
they are a linear function of $\lambda$ with 
matrix coefficients.  Let us express these linear 
functions as $- A_{\bar{z}_1} + \lambda A_{z_2}$ 
and $- A_{\bar{z}_2} - \lambda A_{z_1}$.   
The coefficients $A_{z_1},A_{z_2},A_{\bar{z}_1}, 
A_{\bar{z}_2}$ are to be identified with the 
gauge potentials.  Thus $\Psi(\lambda)$ and 
$\hat{\Psi}(\lambda)$ turn out to satisfy 
(\ref{eq:diffeq-Psi-star}), from which the 
$\star$-product ASDYM equations are derived. 

The other part of the forgoing discussion, too, 
can be mostly extended to the noncommutative 
spacetime. For instance, hidden symmetries are again 
generated by the action 
\beq
    p(\lambda) \ \mapsto \ 
    g_L(\lambda) \star p(\lambda) \star g_R(\lambda)^{-1} 
\eeq
of $GL(N,\bbC)$-valued functions $g_L(\lambda)$ 
and $g_R(\lambda)$ of $(\lambda, z_1 - \lambda \bar{z}_2, 
z_2 + \lambda \bar{z}_1)$.  The associated 
infinitesimal symmetries take the same form as 
those for the ordinary ASDYM equations (with, 
of course, the product of spacetime functions 
being replaced by the $\star$-product).  

An essential difference can be seen in the places 
where finite dimensional linear algebra is used.  
A typical example is the Ward Ansatz.  In the 
noncommutative framework, such a linear algebraic 
structure has to be replaced by an infinite 
dimensional counterpart.  As for the Ward Ansatz, 
for instance, we do not know how to extend it to 
the $\star$-product ASDYM equations.

\newsection{Deformed ADHM construction}

\subsection{How to deform ADHM construction}

The ordinary ADHM construction \cite{bib:ADHM,bib:Atiyah} 
of a $U(N)$-instanton solution with instanton number 
$k$ is based on the $2k \times (2k + N)$ matrix-valued 
function 
\beq
    \Delta(z) = 
    \left(\begin{array}{ccc} 
      B_1 + z_1 1 & B_2 + z_2 1 & I \\
      - B_2^\dagger - \bar{z}_2 1 & 
        B_1^\dagger + \bar{z}_1 1 & J^\dagger 
    \end{array}\right) 
\eeq
of $z = (z_1, z_2, \bar{z}_1, \bar{z}_2)$.  
Here $B_1$ and $B_2$ are $k \times k$ matrices, 
$I$ a $k \times N$ matrix, $J$ an $N \times k$ matrix, 
and $B_1^\dagger, B_2^\dagger, I^\dagger, J^\dagger$ 
their Hermitian conjugate.  Assuming a nondegeneracy 
condition, one can construct a $(2k + N) \times N$ 
matrix $v(z)$ that satisfies the equations 
\beq
    \Delta(z) v(z) = 0, \quad 
    v(z)^\dagger v(z) = 1. 
\eeq
If the so called ADHM equations 
\beq
    [B_1, B_2] + IJ = 0, 
    \nonumber \\ {} 
    [B_1, B_1^\dagger] + [B_2, B_2^\dagger] 
      + I I^\dagger - J^\dagger J = 0 
\eeq
are satisfied, the gauge potentials defined by 
\beq
    A = v(z)^\dagger dv(z) 
\eeq
give a solution (instanton solution) of the ASDYM 
equations. 

As Nekrasov and Schwarz \cite{bib:Ne-Sc} pointed out, 
the instanton solutions of the ASDYM equations on the 
noncommutative $\bbR^4$ can be obtained by deforming 
the ADHM equations as 
\beq
     [B_1, B_2] + IJ = \zeta_\bbC 1, 
     \nonumber \\ {} 
     [B_1, B_1^\dagger] + [B_2, B_2^\dagger] 
      + I I^\dagger - J^\dagger J = \zeta_\bbR 1. 
\eeq
The connection form is now given by the $\star$-product 
\beq
    A = v(z)^\dagger \star dv(z). 
\eeq

\subsection{Solution of Riemann-Hilbert problem} 

We here present, as an application of the ADHM 
construction, an explicit construction of 
the solution of the Riemann-Hilbert problem 
for the instanton solutions.  
This is based on the work of Corrigan et al. 
\cite{bib:Co-Fa-Go-Te} on the Dirac equation 
with the instanton gauge potentials 

According to one of their results, the parallel 
translation (i.e., the ``Wilson line operator'') 
\beq
    w(z,z') = \Pexp \left(\int_{z'}^z A\right) 
\eeq
between two points $z,z'$ on the same twistor 
surface $S(\lambda,u_1,u_2)$  is given by the 
simple formula
\beq
    w(z,z') = v(z)^\dagger v(z'). 
\eeq
Consequently, this matrix obeys the group law 
\beq
    w(z,z') w(z',z'') = w(z, z'') 
\eeq
for any triple $z,z',z''$ of points on 
$S(\lambda,u_1,u_2)$.  

Let us apply this group law to the special 
points 
\beq
    z^\infty(\lambda) 
    &=& (z_1 - \lambda \bar{z}_2, 
        z_2 + \lambda \bar{z}_1, 0, 0), 
    \nonumber \\
    z^0(\lambda) 
    &=& (0, 0, \bar{z}_1 + \lambda^{-1} z_2, 
        \bar{z}_2 - \lambda^{-1} z_1) 
\eeq
that are on the same twistor surface as 
$z = (z_1, z_2, \bar{z}_1, \bar{z}_2)$. 
Acccordingly, we have the relation 
\beq
    w(z^\infty(\lambda),z^0(\lambda)) 
    = w(z, z^\infty(\lambda))^{-1} w(z, z^0(\lambda)). 
\eeq    
This relation is exactly the Riemann-Hilbert problem 
with the patching function 
\beq
    p(\lambda) = w(z^\infty(\lambda),z^0(\lambda)), 
\eeq
for which we thus obtain the explicit solution 
\beq
    \Psi(\lambda) = w(z,z^0(\lambda)), \quad 
    \hat{\Psi}(\lambda) = w(z,z^\infty(\lambda)). 
\eeq

This construction carries over to the noncommutative 
case if the ordinary matrix products therein are 
replaced by the $\star$-product.  The parallel 
translation along the twistor surface is given by 
the $\star$-product 
\beq
    w(z, z') = v(z)^\dagger \star v(z'), 
\eeq
and the foregoing expressions of $g(\lambda)$, 
$\Psi(\lambda)$ and $\hat{\Psi}(\lambda)$ 
remain valid.

\subsection{Remarks on ADHM equations} 

It is well known that the left hand side of the 
ADHM equations, i.e., 
\beq
    \mu_\bbC &=& [B_1, B_2] + IJ, 
    \nonumber \\ 
    \mu_\bbR &=& [B_1, B_1^\dagger] + [B_2, B_2^\dagger] 
      + I I^\dagger - J^\dagger J, 
\eeq
are a pair of moment maps for the hyper-K\"ahler 
quotient construction \cite{bib:Hi-Ka-Li-Ro} of 
the moduli space of (both undeformed and deformed) 
ADHM instantons.  In particular, the pair 
$(\mu_\bbC,\mu_\bbR)$ transforms just like 
the three-vector $(\zeta_\bbC,\zeta_\bbR)$ 
under the $SU(2)$ rotation (\ref{eq:su2-action}) 
of spacetime coordinates.  Therefore it is  natural 
to combine the three moment maps 
$\mu_\bbC,\mu_\bbR,\mu_\bbR^\dagger$ 
to the one-parameter family 
\beq
    \mu(\lambda) 
    = \mu_\bbC + \lambda \mu_\bbR 
      - \lambda^2 \mu_\bbC^\dagger 
    = [B_1 - \lambda B_2^\dagger, B_2 + \lambda B_1^\dagger] 
      + (I - \lambda J^\dagger)(J + \lambda I^\dagger) 
\eeq
of moment maps.  The $SU(2)$ action is now 
represented by fractional transformations 
of $\lambda$: 
\beq
    \lambda \ \longmapsto \
    \frac{- \beta + \alpha\lambda}
         {\bar{\alpha} + \bar{\beta}\lambda}. 
\eeq

A ``pencil'' of moment maps of this type generally 
appears in the quotient construction of the twistor 
space associated with a hyper-K\"ahler quotient 
\cite{bib:Hi-Ka-Li-Ro}. The twistor space $\calZ$ 
of a hyper-K\"ahler manifold is fibered over 
$\bbP_\bbC^1$ by a map $\pi:\calZ \to \bbP_\bbC^1$, 
and each fiber $\pi^{-1}(\lambda)$ is a complex 
symplectic manifold.  The moment map $\mu(\lambda)$ 
is used to make the symplectic quotient of 
$\pi^{-1}(\lambda)$.  Roughly speaking, 
this fiberwise symplectic quotient of $\calZ$ 
gives the twistor space for the hyper-K\"ahler quotient.  

This pencil of moment maps is also interesting 
in the context of finite dimensional integrable 
systems.  Following Gorsky, Nekrasov and Rubtsov 
\cite{bib:Go-Ne-Ru}, let us introduce the 
symplectic form 
\beq
    \Omega = \Tr(dB_1 \wedge dB_2 + dI \wedge dJ) 
\eeq
on the space of the quadruples $(B_1,B_2,I,J)$.  
As they pointed out, $\mu_\bbC$ may be interpreted 
as the moment map of the action 
\beq
    (B_1, B_2, I, J) \ \longmapsto \ 
    (gB_1g^{-1}, gB_2g^{-1}, gI, Jg^{-1}) 
\eeq
of $G = GL(N,\bbC)$, and the reduced phase space 
(actually, with $I$ and $J$ being further constrained 
to a special $G$-orbit) has the structure of an 
integrable system with the Poisson-commutative 
Hamiltonians $\Tr B_2^\ell$, $\ell = 1, \cdots,N$.  
If $k$ is equal to $1$ and  $B_1$ and $B_2$ are 
restricted to Hermitian matrices, this integrable 
system reduces to the rational Calogero-Moser system; 
the case for $k > 1$ is relate to a generalized 
Calogero-Moser system \cite{bib:Gi-He}. 
Now, what occurs if one repeats the same construction 
for the pencil $\mu(\lambda)$ of moment maps?  
Note that the symplectic form, too, has to be 
deformed as 
\beq
    \Omega(\lambda) 
    = \Tr\Bigl(d(B_1 - \lambda B_2^\dagger) 
        \wedge d(B_2 + \lambda B_1^\dagger) 
             + d(I - \lambda J^\dagger)
        \wedge d(J + \lambda I^\dagger)\Bigr). 
\eeq
Upon taking the symplectic quotient, 
a one-parameter family of integrable systems 
will emerge. In fact, $\Omega(\lambda)$ is 
exactly the symplectic form of the fiber 
$\pi^{-1}(\lambda) $ of the twistor space 
before taking the quotient.  Thus the phase space 
of the aforementioned one-parameter family of 
integrable systems turns out to be nothing but 
the fibers $\pi^{-1}(\lambda)$ of the twistor space 
of the instanton moduli space.

\newsection{Two-dimensional reductions}

\subsection{PCF model and Hitchin's equations}

We here examine the PCF model and Hitchin's 
Higgs pair equations as dimensional reductions of 
the ASDYM equations.  

The PCF model can be derived by letting 
\beq
    \nabla_{\bar{z}_1} \to \rd_z + A_z, &&
    \nabla_{z_2} \to \rd_z, 
    \nonumber \\
    \nabla_{\bar{z}_2} \to \rd_{\bar{z}} + A_{\bar{z}}, &&
    \nabla_{z_1} \to \rd_{\bar{z}} 
\eeq
under the gauge $A_{z_1} = A_{z_2} = 0$.  
The associated linear system reads
\beq
    ((1 - \lambda) \rd_z - A_z)
    \Psi(\lambda) = 0, 
    \quad 
    ((1 + \lambda) \rd_{\bar{z}} + A_{\bar{z}})
    \Psi(\lambda) = 0.  
\eeq
On the noncommutative spacetime, $z$ and $\bar{z}$ 
are assumed to obey the commutation relation 
\beq
    [z, \bar{z}] = - \zeta 1 
\eeq
for a real constant $\zeta$, and the linear system 
is replaced by the $\star$-product analogue 
\beq
    ((1 - \lambda) \rd_z - A_z)
    \star \Psi(\lambda) = 0, 
    \quad 
    ((1 + \lambda) \rd_{\bar{z}} + A_{\bar{z}})
    \star \Psi(\lambda) = 0. 
\eeq
Conservation lows, infinitesimal symmetries, 
the Riemann-Hilbert problem, etc. \cite{bib:PCF} 
can be extended to the $\star$-product PCF model 
straightforward.  

Hitchin's Higgs pair equations 
\beq
    F_{z\bar{z}} = [\Phi,\Phi^\dagger], \quad 
    \nabla_{\bar{z}}\Phi = 0, \quad 
    \nabla_z\Phi^\dagger = 0 
\eeq
can be derived from the ASDYM equations by first 
exchanging $z_2 \leftrightarrow \bar{z}_2$ 
(which interchanges anti-self-duality and 
self-duality), then reducing 
\beq
    \nabla_{\bar{z}_2} \to \nabla_z, && 
    \nabla_{z_1} \to \Phi, 
    \nonumber \\
    \nabla_{z_2} \to \nabla_{\bar{z}}, && 
    \nabla_{\bar{z}_1} \to \Phi^\dagger 
\eeq
while letting $\rd_{z_1} \to 0$ and 
$\rd_{\bar{z}_1} \to 0$.  The associated linear 
system can be written 
\beq
    (\nabla_z + \lambda \Phi)\Psi(\lambda) = 0, 
    \quad 
    (\nabla_{\bar{z}} - \lambda^{-1} \Phi^\dagger) 
    \Psi(\lambda) = 0. 
\eeq
A natural $\star$-product analogue of these equations 
are, of course, 
\beq
    F_{z\bar{z}} = [\Phi,\Phi^\dagger]_\star, \quad 
    \nabla_{\bar{z}}\star\Phi = 0, \quad 
    \nabla_z\star\Phi^\dagger = 0 
\eeq
and 
\beq
    (\nabla_z + \lambda \Phi)\star \Psi(\lambda) = 0, 
    \quad 
    (\nabla_{\bar{z}} - \lambda^{-1} \Phi^\dagger) 
    \star \Psi(\lambda) = 0. 
\eeq

\subsection{Some more remarks on Hitchin's equations}

Hitchin's equations are formulated on any compact 
Riemann surface \cite{bib:Hitchin-higgs}. 
If the genus of the Riemann surface is greater 
than one, the moduli space of ``stable'' 
Higgs pairs is a smooth (but noncompact) 
symplectic manifold with the structure of an 
``algebraically integrable Hamiltonian system'' 
\cite{bib:Hitchin-is}.  (This fact is further 
extended to punctured Riemann surfaces including tori.)  
The ``spectral curve'' $\det(\Phi - \zeta 1) = 0$ 
plays a central role therein.  

What about the $\star$-product analogue of Hitchin's 
equations?  Unfortunately, we do not know if the 
moduli space of solutions has any structure of an 
integrable system, because, first of all, the notion 
of determinant (hence, of spectral curve) ceases 
to exist.  This is a place where a linear algebraic 
structure breaks down again.  One can nevertheless 
expect that some yet unknown mechanism might give 
rise to an integrable structure in the moduli space 
solutions.  This issue will be closely related to 
the notion of ``noncommutative Riemann surfaces'' 
that has been pursued by Bertoldi et al. 
\cite{bib:Be-Is-Ma-Pa}. 

Let us finally mention that Hitchin's equations 
are also related to a class of conformal field 
theories --- e.g., the (non-affine) Toda field 
theories \cite{bib:Al-Fa} and $W$-gravity 
\cite{bib:dBo-Go,bib:Go-Ja}.  The associated 
$\star$-product analogues will be interesting 
from the point of view of the Chern-Simons 
and WZW models on noncommutative spaces 
\cite{bib:Mo-Sc,bib:Chu,bib:Da-Kr-La,bib:Fu-In,bib:Nu-Ol-Sc}. 
Note, however, that the naive substitution prescription 
$e^{\alpha \cdot \phi} \to (e^{\alpha \cdot \phi})_\star$ 
in the Toda field theories {\it does not} lead to an 
integrable system.  A correct integrable deformation 
is the so called  ``nonabelian Toda field theory'', 
which does not take such an exponential form.

\newsection{Conclusion}

We have shown that many properties of the 
ASDYM equations are inherited by its analogue on 
the noncommutative $\bbR^4$.  After all, the rule 
of game is quite simple --- just to replace the 
ordinary product by the $\star$-product.  This 
rather naive prescription has turned out to fit 
surprisingly well into the twistorial and integrable 
structures of the ASDYM equations.  Moreover, 
these structures are preserved under dimensional 
reduction to the PCF model and Hitchin's Higgs pair 
equations.  However, linear algebraic structures, 
such as the Ward Ansatz solutions, mostly loose its 
meaning in the noncommutative spacetimes.  

We have also pointed out a few interesting 
structures in the ADHM construction.  These 
structures deserve to be studied in more detail.  

Another important issue, which we have not 
addressed in this paper, is that of the Nahm 
equations.  The Nahm construction of BPS 
monopoles \cite{bib:Nahm,bib:Hitchin-nahm,
bib:Co-Go} has been extended to a noncommutative 
spacetime \cite{bib:Bak}, in which a $\star$-product 
analogue of the Nahm equations is used.   
The $\star$-product Nahm equations have been 
independently studied in the context of the 
M-theory as well \cite{bib:Cu-Fa-Za,bib:Fairlie,
bib:Li-Fa}. 

\subsection*{Acknowledgements}
I am grateful to David Fairlie for useful comments.



\newpage
\begin{thebibliography}{99}

\bibitem{bib:Ne-Sc} 
N. Nekrasov, A. Schwarz, 
Instantons on noncommutative $\bbR^4$ and 
$(2,0)$ superconformal six-dimensional theory, 
{\tt hep-th/9802068}, 
Commun. Math. Phys. {\bf 198}, (1998),  689-703. 

\bibitem{bib:Ka-Ku-Or} 
A. Kapustin, A. Kuznetsov and D. Orlov, 
Noncommutative instantons and twistor transform, 
{\tt hep-th/0002193}. 

\bibitem{bib:ADHM}
M.F. Atiyah, N.J. Hitchin, V.G. Drinfeld and Yu.I. Manin, 
Construction of instantons, 
Phys. Lett. {\bf 65A} (1978), 185-187. 

\bibitem{bib:Atiyah}
M.F. Atiyah, 
Geometry of gauge fields 
(Scuola Normale Superiore, Pisa, 1979). 

\bibitem{bib:Ma-Wo}
L.I. Mason and N.M.J. Woodhouse, 
Integrability, selfduality and twistor theory
(Clarendon Press, Oxford, 1996).  

\bibitem{bib:Polyakov}
A.M. Polyakov, 
Compact gauge fields and the infrared catastrophe, 
Phys. Lett. {\bf B59} (1975), 82-84; 
String representations and hidden symmetries 
for gauge fields, 
Phys. Lett. {\bf B82} (1979), 247-250. 

\bibitem{bib:Hitchin-higgs}
N.J. Hitchin, 
The self-duality equations on a Riemann surface, 
Proc. London Math. Soc. {\bf 55} (1987), 
59-126. 

\bibitem{bib:Mo-Sc}
E.F. Moreno and F.A. Schaposnik, 
The Wess-Zumino-Witten term in noncommutative 
tow-dimensional fermion models, 
{\tt hep-th/0002236}. 

\bibitem{bib:Chu}
C.-S. Chu, 
Induced Chern-Simons and WZW action in 
noncommutative spacetime, 
{\tt hep-th/0003007}. 

\bibitem{bib:Da-Kr-La} 
L. Dabrowski, T. Krajewski and G. Landi, 
Some properties of non-linear $\sigma$-models 
in noncommutative geometry, 
{\tt hep-th/0003099}.

\bibitem{bib:Fu-In}
K. Furuta and T. Inami, 
Ultraviolet property of noncommutative 
Wess-Zumino-Witten model, 
{\tt hep-th/0004024}.  

\bibitem{bib:Nu-Ol-Sc}
C. Nunez, K. Olsen and R. Schiappa, 
From Noncommutative Bosonization to S-Duality, 
{\tt hep-th/0005059}. 

\bibitem{bib:Moyal} 
J. Moyal, 
Quantum mechanics as a statistical theory, 
Proc. Camb. Phil. Soc. {\bf 45} (1949), 99-124. 

\bibitem{bib:Yang}
C.N. Yang, 
Condition of self-duality for $SU(2)$ gauge fields 
on Euclidean four-dimensional space, 
Phys. Rev. Lett. {\bf 38} (1977), 1377-1379. 

\bibitem{bib:Donaldson}
S. Donaldson, 
Anti self-dual Yang-Mills connections over 
complex algebraic surfaces and stable vector bundles, 
Proc. London Math. Soc. {\bf 50} (1985), 1-26. 

\bibitem{bib:Na-Sc}
V.P. Nair and J. Schiff, 
Topological gauge theory and twistors, 
Phys. Lett. {\bf B233} (1989), 343-348; 
A K\"ahler-Chern-Simons theory and quantization 
of the moduli of antiselfdual instantons, 
Phys. Lett. {\bf B246} (1990), 423-429; 
K\"ahler-Chern-Simons theory and symmetries of 
antiselfdual equations, 
Nucl. Phys. {\bf B371} (1992), 329-352. 

\bibitem{bib:Leznov}
A.N. Leznov, 
Deformation of algebras and solution of 
self-duality equation, 
J. Math. Phys. {\bf 28} (1987), 2574-2578. 

\bibitem{bib:Parkes}
A. Parkes, 
A cubic action for selfdual Yang-Mills, 
{\tt hep-th/9203074}, 
Phys. Lett. {\bf B286} (1992), 265-270.  

\bibitem{bib:Pe-Ri}
R. Penrose and W. Rindler, 
Spinors and space-time 
(Cambridge UP, Cambridge, 1986). 

\bibitem{bib:Wa-We}
R.S. Ward and R.O. Wells, Jr., 
Twistor geometry and field theory 
(Cambridge UP, Cambridge, 1990). 

\bibitem{bib:Be-Za} 
A.A. Belavin and V.E. Zakharov, 
Yang-Mills equations as inverse scattering problem, 
Phys. Lett. {\bf 73B} (1978), 53-57.

\bibitem{bib:Pohlmeyer}
K. Pohlmeyer, 
On the Lagrangian theory of anti-self-dual fields 
in four-dimensional Euclidean space, 
Commun. Math. Phys. {\bf 72} (1980), 37-47. 

\bibitem{bib:Ch-Pr-Si}
L.-L. Chau, M.K. Prasad and A. Sinha,  
Some aspects of the linear system for 
self-dual Yang-Mills fields, 
Phys. Rev. {\bf D24} (1981), 1574-1580.

\bibitem{bib:Ward}
R.S. Ward, 
On self-dual gauge fields, 
Phys. Lett. {\bf 61A} (1977), 81-82. 

\bibitem{bib:At-Wa}
M.F. Atiyah and R.S. Ward, 
Instantons and algebraic geometry, 
Commun. Math. Phys. {\bf 55} (1977), 117-124.

\bibitem{bib:At-Hi-Si}
M.F. Atiyah, N.J. Hitchin and I.M. Singer, 
Self-duality in four-dimensional 
Riemannian geometry, 
Proc. R. Soc. London {\bf A362} (1978), 425-461. 

\bibitem{bib:Co-Fa-Ya-Go}
E. Corrigan, D.B. Fairlie, R.G. Yates and P. Goddard, 
The construction of self-dual solutions to 
$SU(2)$ gauge theory, 
Commun. Math. Phys. {\bf 58} (1978), 223-240.

\bibitem{bib:Ivanova}
T. Ivanova, 
Moduli space of self-dual gauge fields, 
holomorphic bundles and cohomology sets, 
{\tt math-th/9902015}. 

\bibitem{bib:Ward-ansatz}
R.S. Ward, 
Ans\"atze for self-dual Yang-Mills fields, 
Commun. Math. Phys. {\bf 80} (1981), 563-574. 

\bibitem{bib:Iv-Ma-Ne}
J.S. Ivancovich, L.J. Mason and E.T. Newman, 
On the density of the Ward Ans\"atze 
in the space of anti-self-dual Yang-Mills solutions, 
Commun. Math. Phys. {\bf130} (1990), 139-155. 

\bibitem{bib:Ch-Ge-Wu}
L.-L. Chau, M.-L. Ge, and Y.-S. Wu, 
Kac-Moody algebra in the self-dual Yang-Mills equation, 
Phys. Rev. {\bf D25} (1982), 1086-1094. 

\bibitem{bib:Dolan}
L. Dolan, 
A new symmetry group of real self-dual Yang-Mills theory, 
Phys. Lett. {\bf 113B} (1982), 387-390. 

\bibitem{bib:Ue-Na}
K. Ueno and Y. Nakamura, 
Transformation theory for anti-self-dual equations 
and the Riemann-Hilbert problem, 
Phys. Lett. {\bf 109B} (1982), 273-278. 

\bibitem{bib:Wu}
Y.-S. Wu, 
The group theoretical aspects of infinitesimal 
Riemann-Hilbert transform and hidden symmetry, 
Commun. Math. Phys. {\bf 90} (1983), 461-472. 

\bibitem{bib:Takasaki}
K. Takasaki, 
Hierarchy structure in integrable systems of 
gauge fields and underlying Lie algebras, 
Commun. Math. Phys. {\bf 127} (1990), 225-238; 
Integrable systems in gauge theory, K\"ahler 
geometry and super KP hierarchy --- symmetries 
and algebraic point of view, 
Proc. International Congress of Mathematicians 
Kyoto 1990, pp. 1205-1214 
(Springer-Verlag, Tokyo, 1991).  

\bibitem{bib:Popov}
A.D. Popov, 
Self-dual Yang-Mills: symmetries and moduli space, 
{\tt hep-th/9803183}, 
Rev. Math. Phys. {\bf 11} (1999), 1091-1149. 

\bibitem{bib:Co-Fa-Go-Te}
E. Corrigan, D.B. Fairlie, P. Goddard and S. Templeton, 
A Green function for the general self-dual gauge field, 
Nucl. Phys. {\bf B140} (1978), 31-44.

\bibitem{bib:Hi-Ka-Li-Ro} 
N.J. Hitchin, A. Kahlhede, U. Lindstr\"om 
and M. Ro\v{c}ek, 
Hyperk\"ahler metrics and supersymmetry, 
Commun. Math. Phys. {\bf 108} (1987), 535-589. 

\bibitem{bib:Go-Ne-Ru} 
A. Gorsky, N. Nekrasov and V. Rubtsov, 
Hilbert scheme, separated variables, and D-branes, 
{\tt hep-th/9901089}.  

\bibitem{bib:Gi-He}
J. Gibbon and T. Hersen, 
A generalization of the Calogero-Moser system, 
Physica {\bf 11D} (1984), 337-348. 

\bibitem{bib:PCF}
A detailed exposition of these issues, along with 
a large list of references, can be found in: 
\newline
J.H. Schwarz, 
Classical symmetries of some two-dimensional models, 
{\tt hep-th/9503078}, 
Nucl. Phys. {\bf B447} (1995), 137-182.  

\bibitem{bib:Hitchin-is}
N.J. Hitchin, 
Stable bundles and integrable systems, 
Duke Math. J. {\bf 54} (1987), 91-144. 

\bibitem{bib:Be-Is-Ma-Pa}
G. Bertoldi, J.M. Isidro, M. Matone and P. Pasti, 
The concept of noncommutative Riemann surface, 
{\tt hep-th/0003200}. 

\bibitem{bib:Al-Fa}
E. Aldrovandi and G. Falqui, 
Geometry of Higgs and Toda fields on Riemann surfaces, 
{\tt hep-th/9312093}, 
J. Geom. Phys. {\bf 17} (1995), 25-48.

\bibitem{bib:dBo-Go} 
J. de Boer and J. Goeree, 
Covariant W-gravity and its moduli space from 
gauge theory, 
{\tt hep-th/9206098}, 
Nucl. Phys. {\bf B401} (1993), 369-412. 

\bibitem{bib:Go-Ja}
S. Govindarajan and T. Jayaraman, 
A proposal for the geometry of $W_n$-gravity, 
{\tt hep-th/9405156}, 
Phys. Lett. {\bf B345} (1995), 211-219.  

\bibitem{bib:Nahm}
W. Nahm, 
The construction of all self-dual multi-monopoles 
by the ADHM method, 
in: N.S. Craigie, P. Goddard and W. Nahm, (eds.), 
{\it Monopoles in quantum field theory}, pp. 87-94 
(World Scientific, Singapore, 1982). 

\bibitem{bib:Hitchin-nahm}
N.J. Hitchin, 
On the construction of monopoles, 
Commun. Math. Phys. {\bf 89} (1983), 145-190. 

\bibitem{bib:Co-Go}
E. Corrigan and P. Goddard, 
Construction of monopole solutions and reciprocity, 
Ann. Phys. {\bf 154} (1984), 253-279. 

\bibitem{bib:Bak} 
D. Bak, 
Deformed Nahm equation and a noncommutative 
BPS monopoles, 
{\tt hep-th/9910135}. 

\bibitem{bib:Cu-Fa-Za}
T. Curtlight, D.B Fairlie and C.K. Zachos, 
Integrable symplectic trilinear interaction terms 
for matrix membranes, 
{\tt hep-th/9704037}, 
Phys. Lett. {\bf B405} (1997), 37-44. 

\bibitem{bib:Fairlie}
D.B. Fairlie, 
Moyal brackets in M-theory, 
{\tt hep-th/9707190}, 
Mod. Phys. Lett. {\bf A13} (1998), 263-274. 

\bibitem{bib:Li-Fa}
L. Baker and D. Fairlie, 
Moyal Nahm equations, 
{\tt hep-th/9901072}, 
J. Math. Phys. {\bf 40} (1999), 2539-2548. 

\end{thebibliography}
\end{document}